\documentclass[aps,prb,reprint,showpacs,superscriptaddress]{revtex4-1}

\usepackage{color}
\usepackage{graphicx}
\usepackage{amsmath}
\usepackage{amssymb}
\usepackage{dsfont}

\newcommand{\beq}{\begin{eqnarray}}
\newcommand{\eeq}{\end{eqnarray}}

\begin{document}

\title{Bipartite entanglement dynamics of two-level systems in sub-Ohmic reservoirs }

\date{\today}

\author{Denis Kast}
\affiliation{Institut f\"{u}r Theoretische Physik, Universit\"{a}t Ulm,
  Albert-Einstein-Allee 11, 89069 Ulm,
  Germany}
\author{Joachim Ankerhold}
\affiliation{Institut f\"{u}r Theoretische Physik, Universit\"{a}t Ulm,
  Albert-Einstein-Allee 11, 89069 Ulm,
  Germany}

\begin{abstract}
The quantum dynamics of pairs of two level systems immersed in dissipative reservoirs with sub-Ohmic spectral distributions
 is studied by means of numerically exact path integral Monte Carlo methods. It is shown that this class of reservoirs, relevant for generic properties of strongly non-Markovian environments with possible realizations in solid state structures at cryogenic temperatures, supports bipartite entanglement in broad ranges of parameter space and even for broken symmetries and in presence of thermal noise. The sensitivity of  creation of entanglement in non-equilibrium on initial preparations of the reservoir is investigated in detail.
\end{abstract}


\maketitle

\paragraph{Introduction-}
Entanglement is one of the most fascinating properties of many-body quantum systems and
is thus a key resource for quantum information processing \cite{huang}. In any realization, however, the impact of
environmental degrees of freedom has to be taken into account which is particularly true for
solid state implementations in form of e.g.\ circuit quantum electrodynamical settings \cite{makhlin,blais,shumeiko}, superconducting qubits \cite{lofranco2}, or aggregates utilizing  NV-centers in diamonds \cite{jelezko}. Here,
significant progress has been made in extending decoherence times of single and multiple
two level systems (TLS) and in developing protocols for quantum error correction \cite{shumeiko}.

In solid state realizations, high frequency noise is typically of Ohmic type with a mode distribution $I(\omega)\propto \omega$ (up to a cut-off frequency) while in the regime
of lower frequencies non-Ohmic behavior $I(\omega)\propto \omega^s, s\neq 1$ seems to dominate \cite{fnoise,paladino,imamoglu,galland}. Particularly sub-Ohmic reservoirs with $s<1$ have gained substantial
attention recently ranging from fundamental aspects such as dissipation induced quantum phase transitions \cite{anders,rieger,exdiag,thoss2,vojta}, system-reservoir entanglement \cite{hofstetter,chin,wang} and
coherent-incoherent transitions \cite{thorwart,kast,kastprb,thorwart2} to their possible impact in specific realizations e.g.\ mesoscopic rings \cite{ring}, chains of trapped ions \cite{porras}, or excitons in suspended carbon nanotubes \cite{imamoglu,galland}.

Recently, entanglement of bipartite TLS in presence of conventional Ohmic reservoirs has gained increased attention 
in contexts including entanglement dynamics due to individual \cite{bellomo} or common \cite{braun,bose,thorwart1,mccutcheon,makri,marzolino} environments and for non-equilibrium steady states \cite{briegel}. Much less is known though about the impact of non-Ohmic reservoirs associated with strong non-Markovian dynamics \cite{tanimura,plenio} beyond weak coupling treatments \cite{lofranco}. Here, we consider the most challenging low temperature regime for broadband sub-Ohmic reservoirs non-perturbatively  to reveal the interplay of sluggish low and dynamical high frequency modes for the survival and creation of quantum non-locality.

For that purpose, path integral Monte Carlo techniques (PIMC) \cite{lothar1} are employed to treat the
non-equilibrium dynamics of pairs of TLS also at stronger dissipation and zero temperature. A translational invariant
coupling between TLS and surrounding degrees of freedom guarantees that the latter ones act only {\em dynamically}
onto the system and that in contrast to alternative settings \cite{makri,marzolino} any reservoir induced static couplings between the TLS are absent \cite{leggett,weiss}. We show that sub-Ohmic environments preserve certain types of initially prepared entangled states completely even when symmetries (decoherence free subspace) are broken. For initially separable TLS, this class of reservoirs is able to dynamically create entanglement in broad regions of parameter space, thereby sensitively depending on the initial bath state. These findings may be of relevance not only for
 quantum open systems in general but also for applications in solid state structures, where low frequency bath modes prevail at low temperatures  \cite{paladino}.

\paragraph{Model-}

We consider a compound, where a pair of TLS  is immersed in a thermal reservoir \cite{leggett,weiss,breuer} so that $H = H_{\rm TLS} + H_R + H_I$. In a minimal setting, the bare system consists of unbiased TLS, denoted $A$ and $B$ henceforth, with ferromagnetic coupling in $x$-direction, i.e.,
\beq
    H_{\rm TLS} = - \frac{\hbar\Delta_A}{2} \sigma^A_x - \frac{\hbar\Delta_B}{2} \sigma^B_x - \hbar J \sigma_x^A \sigma_x^B\, ,
\label{H_2TLS}
\eeq
with $\sigma_p^{A/B}, p=x, y, z$ denoting Pauli matrices. This system interacts with a reservoir
\beq
H_R+H_I =\!\sum_{\alpha}\!\frac{p_{\alpha}^2}{2m_{\alpha}}+\frac{m_{\alpha} \omega_{\alpha}^2}{2}\left(x_\alpha+ c_\alpha\, \frac{\sigma_z^A+\sigma_z^B}{2 m_\alpha\omega_\alpha^2}\right)^2 	
\label{HISM}
\eeq
 where the TLS-reservoir interaction is assumed to be translational invariant according to typical situations for solid state devices \cite{weiss,makhlin}.
 Thus, the bath acts on the TLS only dynamically, a situation for which, in absence of a direct coupling $J=0$, it is not known to what extent a sub-Ohmic medium may turn into an entanglement generating agent \cite{bose,schmidt}. In the non-translational case, a common bath induces a {\em static} $\sigma_z^A \sigma_z^B$-interaction which easily supports bi-partite entanglement \cite{makri}. Further, in (\ref{HISM}) only the net spin polarization occurs, corresponding e.g.\ to a negligible spatial separation of the TLS \cite{bose}. In the continuum limit, a sub-Ohmic bath is characterized by  a spectral distribution
\beq
   I(\omega) = 2\pi\alpha\, \omega_c^{1-s}\, \omega^s\  {\rm e}^{-\omega/\omega_c}\, ,\label{spectral}
\eeq
with $0<s<1$ and cut-off $\omega_c$. Accordingly, the portion of low frequency modes is enhanced compared to the Ohmic case ($s=1$) \cite{weiss} which gives rise to intricate quantum-classical transitions for single TLS \ \cite{anders,rieger,exdiag,thoss2,hofstetter,chin,wang,thorwart,kast,thorwart2}.

Two basis sets are particularly suitable to represent (\ref{H_2TLS}), namely, the localized basis (eigenstates $|\sigma\rangle_{A/B}$ of $\sigma_z^{A/B}$-operators) and the delocalized (entangled) basis of Bell states. The PIMC is conveniently formulated in
\beq
|\sigma+\sigma', \sigma-\sigma'\rangle = |\sigma\rangle_A |\sigma'\rangle_B\ ; \ \sigma, \sigma'=\pm 1
\label{ISMbasis}
\eeq
which allows to express the Bell-basis as
\beq
   |\Phi_{\pm} \rangle = \frac{|2,0\rangle \pm |-2,0\rangle}{\sqrt{2}},\,
   |\Psi_{\pm} \rangle = \frac{|0,2\rangle \pm |0,-2\rangle}{\sqrt{2}}.
\label{Bell-states}
\eeq
Initially, the pair of TLS can now be prepared in a separable state $\rho_{\rm TLS}(0) = \rho^A(0) \otimes \rho^B(0)$ with an arbitrary orientation of the individual TLS
\beq
\rho^{A/B}(0) &=& \frac{1}{2}\left(\mathds{1}+\sum_{p=x,y,z} \langle \sigma_{p}^{A/B}(0)\rangle\, \sigma_{p}^{A/B}\right)\,
\eeq
to analyze the creation of entanglement. Alternatively, one can start from the Bell-states in (\ref{Bell-states}) to monitor the survival of an initially entangled bi-partite state.

 The expectation value of an observable $\Pi$ follows from
\beq
   P_{\pi}(t) = {\rm Tr}\{\exp\left(iHt/\hbar\right) W(0) \exp\left(-iHt/\hbar\right) \Pi\}\, ,
\label{observable}
\eeq
where $W(0)=\rho_{\rm TLS}(0) W_{\beta, \rm R} $  is a factorized initial state of the compound. It includes a thermal state of the reservoir
\beq
   W_{\beta, \rm R} =\frac{1}{Z} e^{-\beta \left(H_R + \mu \sum_\alpha c_\alpha x_\alpha \right)}\label{inibath}
\eeq
with $\beta=1/k_B T$ and a parameter $\mu$ which allows to tune a displacement of equilibrated bath modes \cite{lucke,weiss}. Physically,
the TLS may initially be held in a fixed spin orientation to which the bath equilibrates so that $\mu=\langle\sigma_z^A(0)+\sigma_z^B(0)\rangle/2$.
In a complementary setting, the TLS is driven out of equilibrium by a short pulse so that the bath modes have no time to rearrange. While for Ohmic reservoirs this is not relevant (see below), the situation is different for sub-Ohmic media with substantial portions of sluggish modes. In fact,  we will show that the initial state of the reservoir may have profound impact on the TLS dynamics. Now, by choosing in (\ref{observable}) as observables a complete set of projectors onto the states (\ref{ISMbasis}), the reduced density is fully characterized.
\begin{figure}
\centering
\includegraphics[angle=270,width=0.85\columnwidth]{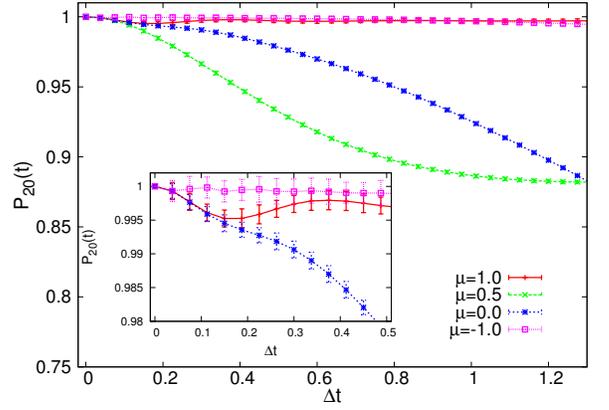}
\vspace*{0.3cm}
\caption{Population dynamics of the localized state $|2,0\rangle$ with $J=0$ in sub-Ohmic baths with $s=0.25$, $\alpha=0.3$, $\omega_c/\Delta=10$ at $T=0$ and for various initial orientations $\mu$.}
\label{s025-osci}
\end{figure}

For the dynamics of TLS in sub-Ohmic reservoirs conventional perturbative treatments fail due to strong non-Markovian effects.
PIMC techniques have been thus developed as numerically exact means to cover the full range from high to low temperatures, weak to strong dissipation, arbitrary TLS initial preparations, and sufficiently long times \cite{lothar1,kast}.
After integrating out the reservoir degrees of freedom observables can be expressed as a double path integral with forward
$\sigma^{A/B}$ and backward $\tilde{\sigma}^{A/B}$ paths
in the basis of the individual $\sigma_z^{A/B}$ operators \cite{weiss}. Switching to the combinations
${\eta}^{A/B}=(\sigma^{A/B}+\tilde{\sigma}^{A/B})/2$ and
${\xi}^{A/B}=(\sigma^{A/B}-\tilde{\sigma}^{A/B})/2$ then yields
\begin{equation}
P_{\pi}(t) = \oint\!{\cal D}\vec{\eta} \oint\!{\cal
D}\vec{\xi}\; \Pi(\vec{\eta},\vec{\xi}) {\cal A}[\vec{\eta},\vec{\xi}]\, \exp\!\left( -
 \Phi[\vec{\eta},\vec{\xi}] \right) \label{pops}\;
\end{equation}
with $\vec{\eta}=(\eta^A,\eta^B)$ and $
\vec{\xi}=(\xi^A,\xi^B)$. Here, ${\cal A}$
is the bare action factor in absence of a reservoir and
\begin{eqnarray}
\Phi[\vec{\eta},\vec{\xi}] &=& \frac{1}{2}\int_0^t dv \int_0^t du
\left\{\left[\vec{\xi}(v) \cdot \vec{e}\right] L'(v-u)
\left[\vec{\xi}(u) \cdot \vec{e}\right]\right.\nonumber\\
&&\left.\hspace{1.5cm} +i \left[\vec{\xi}(v) \cdot
\vec{e}\right] L''(v-u)
\Big[\vec{\eta}(u) \cdot \vec{e}\Big]\right\}\nonumber\\
&& +\frac{i}{2}\int_0^t dv \left[ \vec{\xi}(v) \cdot
\vec{e}\right]\Big\{\gamma(0)
\Big[\vec{\eta}(v) \cdot \vec{e}\Big]-2\mu\gamma(v)\Big\}\nonumber \\
\label{influ}
\end{eqnarray}
with $\vec{e}=(1,1)$ is the influence functional (IF) capturing reservoir induced
self-interactions via the bath correlation \cite{weiss} $L(t)=L'(t)+i L''(t)$, i.e.,
\begin{equation}\label{kernel}
L(t)= \int_0^\infty \frac {d\omega}{\pi} I(\omega) [{\rm coth}(\frac{\omega\hbar\beta}{2})\cos(\omega t)-i\sin(\omega t)].
\end{equation}
The translational invariant coupling (\ref{HISM}) results in a term containing the classical damping kernel $d\gamma(t)/dt=2 L''(t)$ with  $\gamma(0) = 4\alpha\omega_c \Gamma(s)$ for sub-Ohmic reservoirs. The $\mu$-dependence describes the bath preparation (\ref{inibath}).

\paragraph{Reservoir-induced coherent dynamics-}
Single TLS in Ohmic-type of reservoirs, i.e.\ $s=1$ in (\ref{spectral}) with large $\omega_c$,  relax at $T=0$ to thermal equilibrium either via damped oscillations ($\alpha<1/2$, coherent) or via classical-like monotonous decay ($\alpha>1/2$, incoherent). This is not the case in sub-Ohmic reservoirs for spectral exponents $s<0.5$, where coherences are preserved for arbitrary coupling \cite{kast}. Here, we focus on the dynamics of pairs of TLS depending on the initial preparation of the bath (\ref{inibath}). This will be  of relevance for the creation of entanglement (see below).

A typical example is shown in Fig.~\ref{s025-osci} with the population of the localized state $|2,0\rangle$ for various bath preparations $\mu$. For $\mu=\pm 1$ the population weakly oscillates close to its initial value while substantial deviations are seen for $\mu= 0, 0.5$. To obtain an understanding of these features, we consider the IF (\ref{influ}). For $\mu\neq 0$ the bath preparation part acts like a time dependent bias \cite{weiss}  corresponding to a $(\sigma_z^A+\sigma_z^B)$ term in $H_{\rm TLS}$ (\ref{H_2TLS}). In the short time range and for $s\ll 1$, this bias is basically static and large $\gamma(t)\approx \gamma(0)= 4\alpha\omega_c/s\gg J, \Delta_{A/B}$, which forces the density to be almost diagonal with $[\vec{\eta}(v) \cdot \vec{e}][\vec{\xi}(v) \cdot \vec{e}]\approx \vec{\xi}(v) \cdot \vec{e}$. Hence, due to $L''(0)=0$
\begin{equation}\label{phired}
\Phi\approx \frac{i}{2}\int_0^t dv \left[ \vec{\xi}(v) \cdot
\vec{e}\right] \gamma(0) (1-2\mu)\,
\end{equation}
the IF turns into an effective net bias $\epsilon\approx \gamma(0) (1-2\mu)/2$ for both TLS.  For $\mu\neq 0$ and $ 1-2\mu\neq 0$, this leads to a trapping of initially localized states superposed with weak high frequency oscillations of order $|\epsilon|$ as seen in the PIMC data. For $\mu=0.5$, the effective bias is absent and the TLS tends to follow its bare dynamics for short times. The above reasoning, however, does not apply to the preparation $\mu=0$. In this case, the initial bath state does not induce a nearly diagonal density, decoherence sets in immediately and leads to a decaying dynamics (cf.~Fig.~\ref{s025-osci}). Note that for a purely Ohmic bath $\gamma(t)\propto \delta(t)$ any bath preparation drops out for localized initial states. Further,  a non-translational invariant system-bath coupling corresponds to a bias $\epsilon_{\rm non}=-\mu\gamma(0)$ with no symmetry around $\mu=0.5$.

\begin{figure}
\centering
\includegraphics[width=0.85\columnwidth]{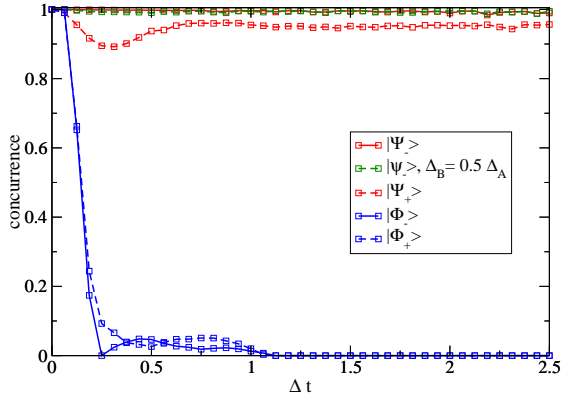}
\caption{Survival of bi-partite entanglement starting from Bell state preparations (\ref{Bell-states}) in presence of a sub-Ohmic reservoir ($s=0.25, T=0, \alpha=0.1$). In addition to the symmetric situation $\Delta_A=\Delta_B$ (blue, red) with $J=0$, the case $\Delta_A\neq \Delta_B$ (green) is shown.}
\label{Bellfig}
\end{figure}
\paragraph{Entanglement dynamics-}
 We now analyze bi-partite entanglement due to the dynamical influence of sub-Ohmic reservoirs in the deep quantum regime. A reliable measure  is the concurrence \cite{concurrence} $C(\rho) := {\rm max}(0, \lambda_1 - \lambda_2 - \lambda_3 - \lambda_4)$, where $\lambda_i$ are the eigenvalues of the matrix $R=(\rho^{1/2} \tilde{\rho} \rho^{1/2})^{1/2}$
sorted in descending order with $\tilde{\rho} = (\sigma_y^A \otimes \sigma_y^B) \rho^* (\sigma_y^A \otimes \sigma_y^B)$.

First, the dynamics of the concurrence for initially maximally entangled states $|\Phi_\pm\rangle$ and $|\Psi_\pm\rangle$  (\ref{Bell-states}) with $C=1$ is studied (see Fig.~\ref{Bellfig}). As shown previously, the symmetry of the system-bath coupling in (\ref{H_2TLS}) supports the existence of a decoherence free subspace (DFS) \cite{lidar}. Namely, the Bell states $|\Psi_{\pm}\rangle$ are eigenstates of the coupling operator $\sigma_z^A+\sigma_z^B$ with zero eigenvalue. In addition, in the symmetric case $\Delta_A=\Delta_B$, the state $|\Psi_-\rangle$ is an eigenstate of $H_{TLS}$  with zero eigenvalue  meaning that under these conditions this state spans a one-dimensional DFS.
Thus, if the TLS is prepared in a pure state $|\Psi_-\rangle$, the initial concurrence $C=1$ is preserved during the time evolution.
If symmetry is broken $\Delta_A\neq \Delta_B$,  the concurrence is expected to degrade due to a process, where first the bare dynamics populates states $|\Phi_\pm\rangle$ which are subject to decoherence. Notably, the PIMC results reveal that sub-Ohmic baths support entanglement initially stored in $|\Psi_-\rangle$ even for substantial asymmetry in contrast to Ohmic dissipation (not shown). A similar scenario applies for an initial preparation in the non-dissipative state $|\Psi_+\rangle$, but is not the case for the Bell states $|\Phi_\pm\rangle$. The latter experience a sharp drop of quantum non-locality towards $C=0$, a phenomenon known as sudden death of entanglement \cite{lofranco}. We mention that the DFS is also broken for a non-symmetric system-bath coupling as previously studied for Ohmic dissipation \cite{bose}.
\begin{figure}
\centering
\includegraphics[width=0.85\columnwidth]{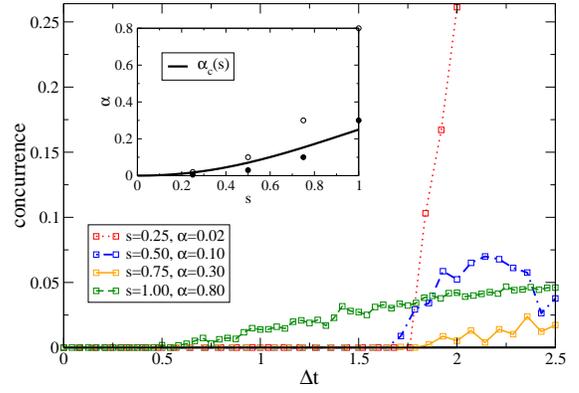}
	\caption{Concurrence of a symmetric TLS with $J=0$ starting from the separable state $|2,0\rangle$ and  sub-Ohmic reservoirs with initial preparation $\mu=0$ for parameters above (inset, open dots)  the threshold (\ref{threshold}) at $T=0$. Data for $C=0$ (inset, filled dots: $s=0.25, \alpha=0.005$; $s=0.5, \alpha=0.03$; $s=0.75, \alpha=0.1$, $s=1, \alpha=0.3$) are not included.}
\label{alphagfig}
\end{figure}

 Let us now turn to localized initial state preparations (separable initial states) according to (\ref{ISMbasis}) and monitor to what extent entanglement is {\em dynamically} created.  Analytical results are available in limiting cases for the symmetric situation $\Delta\equiv\Delta_A=\Delta_B$. In thermal equilibrium with negligible system-bath coupling, one finds \cite{tanimura}
\begin{equation}
C_\beta={\rm max}\left[0, \frac{{\rm sinh}(\hbar\beta J)-1}{{\rm cosh}(\hbar\beta\Delta)-{\rm cosh}(\hbar\beta J)}\right]\, ,
\end{equation}
 while the bare dynamics (no bath) leads to
 \begin{equation}\label{conbaretls}
 C_{\rm TLS}(t)=|\sin(2 J t)|\, .
 \end{equation}
In both cases, any vanishing ferromagnetic coupling $J=0$ implies always separable states $C=0$.

Thus, for the PIMC simulations we begin with the challenging case $J=0$ where only the common reservoir may act as an agent to induce entanglement. Formally, this can be read off the IF (\ref{influ}) which in (\ref{pops}) favors spin orientations with $\vec{\xi}\cdot \vec{e}\approx 0$: Those with $\xi^A=\xi^B=0$ induce decoherence (diagonal density) while those with $\xi^A=-\xi^B\neq 0$ create entanglement, i.e.\ overlap
 with Bell states $|\Psi_\pm\rangle$ (\ref{Bell-states}). Whether initially separable states evolve into entangled ones, depends on which of these orientations are favored dynamically. Hence, according to the discussion around Fig.~\ref{s025-osci}, we put $\mu=0$ to avoid any reservoir induced trapping. Then, to roughly identify regions in parameter space where entanglement can be expected, we argue as follows: The $\vec{\eta}$-dependent part of the IF supports also overlap with Bell states $|\Phi_\pm\rangle$ for $\vec{\eta} \cdot \vec{e}\approx 0$ where $\eta_A\neq 0$. If this part dominates, entanglement creating spin orientations may be privileged against decoherence inducing ones.  To compare the respective terms in the IF (\ref{influ}), we estimate the size of $\vec{\xi}\cdot \vec{e}$ to be at most of order $1/\sqrt{L'(0)/2}$. Next, a partial integration with $2 L''(t)=d\gamma(t)/dt$ cancels the $\gamma(0)$-part and produces a term with $\frac{1}{2}[\vec{\xi}(v)\cdot \vec{e}] \gamma(v-u) d[\vec{\eta}(u) \cdot \vec{e}]/du$. Thus, we require $\gamma(0)/\sqrt{2 L'(0)}>1$ so that for $T=0$
\begin{equation}\label{threshold}
4\alpha\, \frac{\Gamma(s)^2}{\Gamma(1+s)}>1\, .
\end{equation}
 This then defines $\alpha_C(s)=\Gamma(1+s)/[4 \Gamma(s)^2]$ {\em below} which reservoirs are not expected to support entanglement. If it appears, it does so on the time scale $2/\sqrt{\gamma(0)}$.

In  Fig.~\ref{alphagfig} PIMC data for the concurrence are shown for various coupling parameters and spectral exponents. The results are indeed in agreement with (\ref{threshold}) which is remarkable given the rough estimates on which it is based. Finite concurrence is not seen for $\alpha<\alpha_C(s)$ with e.g.\ $\alpha_C(0.25)\approx 0.017$, $\alpha_C(0.5)\approx 0.07$,  $\alpha_C(0.75)\approx 0.15$, $\alpha_C(1)=0.25$. For $\alpha>\alpha_C(s)$ we find $C>0$ only  in the range $s\lesssim 0.8$, while towards the Ohmic regime, $\alpha_C$ sets only a lower bound due to the growing portion of high frequency bath modes. The condition (\ref{threshold}) thus provides an understanding of why deeper in the sub-Ohmic regime entanglement is created for weaker dissipation. In contrast, a non-translational coupling in (\ref{HISM}) promotes entanglement in the Ohmic regime even for weak couplings\cite{makri,marzolino}.

\begin{figure}
\centering
\includegraphics[width=0.85\columnwidth]{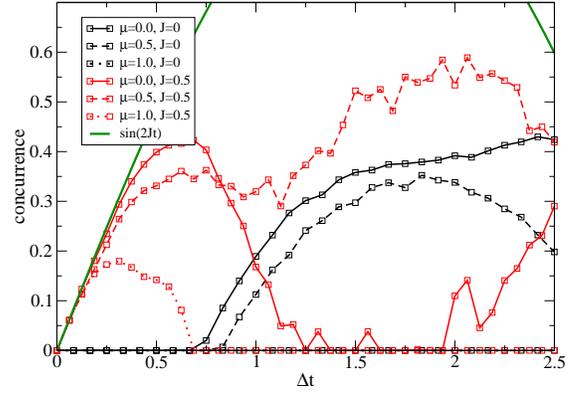}
\caption{Concurrence in a sub-Ohmic bath with $s=0.25$ at $T=0$ starting from various initial bath orientations $\mu$. Shown are results without ($J=0$) and with direct coupling of symmetric TLS. Other settings are as in Fig.~\ref{alphagfig}.}
\label{entang_mu}
\end{figure}
\begin{figure}
\centering
\includegraphics[width=0.85\columnwidth]{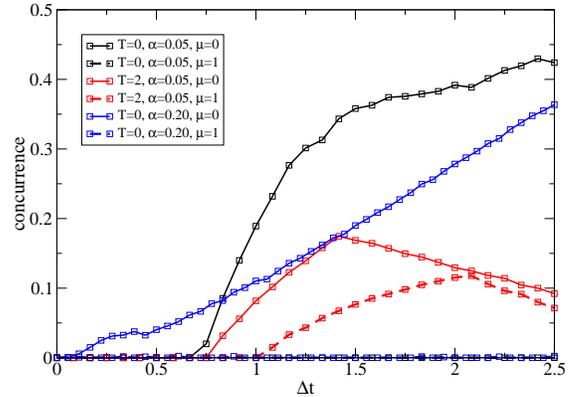}
\caption{Same as in Fig.~\ref{entang_mu} for $J=0$ but also at finite temperatures $T\neq 0$.}
\label{entang_T}
\end{figure}
Next, we discuss the dependence of entanglement creation on the initial bath preparation and the direct TLS coupling (cf.~Fig.~\ref{entang_mu}).
Initially,  the TLS is again  prepared in the separable state $|2,0\rangle$ to which the bath is equilibrated for $\mu=1$.
In case of uncoupled TLS, for this latter initial bath state entanglement is not induced on the time scale of the PIMC, in contrast to substantial concurrence for preparations with $\mu=0$ and $\mu=0.5$. This behavior can be attributed to what we discussed above (see Fig.~\ref{s025-osci}): a large effective bias induced by low frequency modes of the bath basically traps any initially localized state, thus suppressing any dynamical production of entanglement.
If the TLS are directly coupled $J\neq 0$, the concurrence starts to follow initially the bare result (\ref{conbaretls}) for all $\mu$. However, for $\mu=1$ this seems to be only a transient phenomenon with small amplitude, while in other cases amplitudes are larger and even revivals are found ($\mu=0$).

In recent studies for Ohmic dissipation \cite{bose,xu}, thermal fluctuations have always been detrimental to the survival of entanglement. On the other hand, it is known that non-local quantum correlations may be promoted by finite noise, see e.g.\ \cite{caruso,plenio}. In  Fig.~\ref{entang_T} we present PIMC data for the concurrence at finite temperatures in comparison. Indeed, for $\mu=0$ thermal noise  suppresses entanglement while the opposite is true for  $\mu=1$, where $C=0$ at $T=0$. There, fluctuations seem to weaken the trapping effect of the bath to the extent that they play a constructive role in the creation of a finite concurrence. Notably, this even occurs at temperatures which exceed the bare level spacing of the TLS.

\paragraph{Conclusion-}Bi-partite entanglement dynamics has been investigated by means of numerically exact PIMC simulations. The survival and creation of entanglement in presence of common sub-Ohmic reservoirs has been analyzed. The relatively large portion of low frequency modes in this class of reservoirs supports substantial entanglement
in broad ranges of parameter space and sensitively depends on the initial state of the reservoir. The phenomenological model considered here reveals generic features which may be of relevance not only from a fundamental point of view but also for solid state devices at cryogenic temperatures and in the context of reservoir engineering.

\acknowledgments{We thank J.T. Stockburger for valuable discussions. Financial support from the DFG through the SFB/TRR21 and the GIF is gratefully acknowledged.}

\end{document}